\documentclass[sigconf]{acmart}
\usepackage{color}
\usepackage{fontawesome5}
\newcommand{\n}{{\sc PortfolioMentor} }
\AtBeginDocument{%
  \providecommand\BibTeX{{%
    \normalfont B\kern-0.5em{\scshape i\kern-0.25em b}\kern-0.8em\TeX}}}
\setcopyright{acmcopyright}
\copyrightyear{2023}
\acmYear{2023}
\acmDOI{}
\acmConference[]{UIST '23 SIC}{October 29 – November 1, 2023}{San Francisco, CA}
\acmPrice{15.00}
\acmISBN{978-1-4503-XXXX-X/18/06}
\begin{document}

\title[{PortfolioMentor}: Multimodal Generative AI Companion for\\Learning and Crafting Interactive Digital Art Portfolios]{{PortfolioMentor}: Multimodal Generative AI Companion for Learning and Crafting Interactive Digital Art Portfolios}
\author{Tao Long}
\authornote{Both authors contributed equally to this research.\vspace{-1px}}
\email{long@cs.columbia.edu}
\affiliation{
  \institution{Columbia University}
  \city{New York}
  \state{New York}
  \country{USA}
}

\author{Weirui Peng}
\authornotemark[1]
\email{wp2297@columbia.edu}
\affiliation{
  \institution{Columbia University}
  \city{New York}
  \state{New York}
  \country{USA}
}

\renewcommand{\shortauthors}{Long and Peng}

\begin{abstract}
Digital art portfolios serve as impactful mediums for artists to convey their visions, weaving together visuals, audio, interactions, and narratives.
However, without technical backgrounds, design students often find it challenging to translate creative ideas into tangible codes and designs, given the lack of tailored resources for the non-technical, academic support in art schools, and a comprehensive guiding tool throughout the mentally demanding process.
Recognizing the role of companionship in code learning and leveraging generative AI models' capabilities in supporting creative tasks, we present \n, a coding companion chatbot for IDEs. This tool guides and collaborates with students through proactive suggestions and responsible Q\&As for learning, inspiration, and support.
In detail, the system starts with the understanding of the task and artist's visions, follows the co-creation of visual illustrations, audio or music suggestions and files, click-scroll effects for interactions, and creative vision conceptualization, and finally synthesizes these facets into a polished interactive digital portfolio.
\end{abstract}


\maketitle

\begin{figure}[h]
  \centering
  \includegraphics[width=\linewidth]{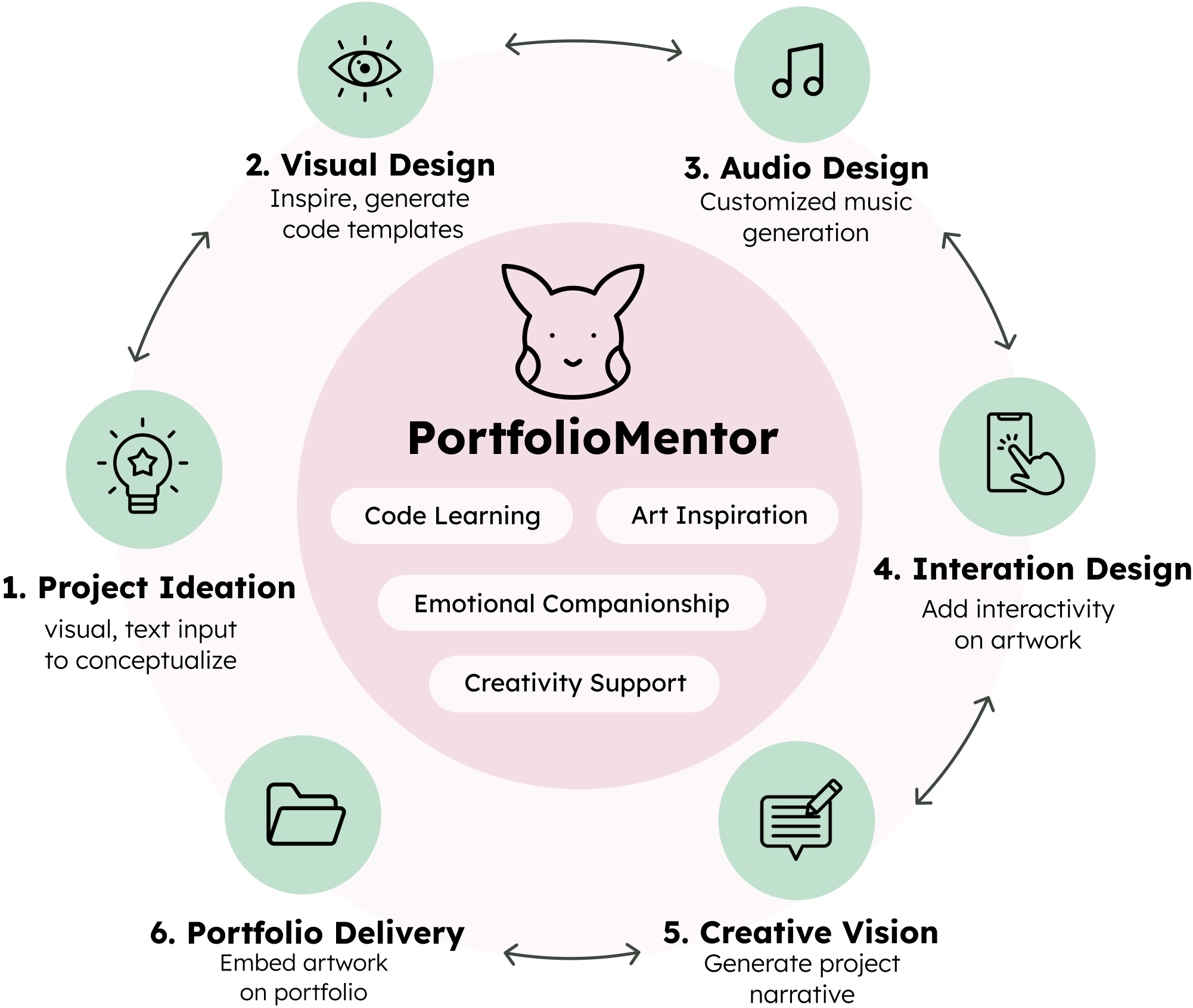}
  \caption{\n System Affordances}
    \label{fig:taxonomy}
  \Description{Our system provides six affordances, including project initiation, isual design, audio design, interaction design, creative vision narrtives, and final delivery.}
  \vspace{-20px}
\end{figure}

\section{Introduction}
Digital art portfolios serve as an impactful medium for artists to convey their creative visions, weaving dimensions like visuals, audio, and interactions together~\cite{creativevision, multimediaportfolio}. However, crafting an interactive artwork from scratch is always tedious and undersupported through its iterations, trials, and errors~\cite{classroomCHIcreativecoding, creativecoderneeds}. Particularly, early-stage design students, often without technical backgrounds, encounter obstacles. From our formative study (N=5), both students and instructors shared it was challenging to translate creative ideas into tangible designs and codes, given the lack of academic resources, support, and a need for companionship throughout the process.

Generative models raise huge potential in supporting the brainstorming and exploration process for making digital art, given the domain knowledge and contextual understanding capabilities in creative tasks. For example, various generative AI-based tools have been used to extend design professionals' creativity, covering visual and interface design~\cite{promptinfuser,stylette,creativecodingexploration}, audio and video design~\cite{disco,NoviceAImusic}. 

In response, we designed \n, an interactive coding companion tool for non-technical design students. Recognizing AI's companionship and availability inside the coding learning and creativity support process \cite{socialDynamics, companionshipcoding, companionshipcreative}, \n guides and collaborates with the students with proactive coding and design suggestions and responsible Q\&A. Similar to ~\cite{Tweetorial_hook, scaffolding,reelframer}, this tool aims to scaffold and alleviate the tedious and mentally demanding process of building an interactive art portfolio through comprehensive creativity and coding support. Our system incorporates multiple affordances (See Figure \ref{fig:taxonomy}) to foster learning and encourage the development of unique creative ideas beyond visual presentations. Based on various multimodal generative models, \n supports the co-creation process of digital artwork including visuals, animations, background sounds, interaction designs, creative visions and narratives, and deployable code snippets.

\section{Formative Study \& Design Goals}
We conducted semi-structured interviews with four design students (S1-S4), each pursuing a degree in fine art or visual design, and one instructor (I1) with over ten years of teaching experience in coding for art students. Each interview lasted 30 minutes, exploring personal experiences and identifying coding-related obstacles. Here are four challenges they faced in the process of building an interactive art portfolio, following with design goals for \n:

\noindent\textbf{\\1. Information Barrier.}
All participants highlighted that there exists a significant information barrier for non-technical students while navigating code-learning materials and forums online.\\
\textbf{{$\bigstar$} Design Goals:} \n should incorporate LLMs and visual models to retarget technical content into natural, simplified, and visual languages for non-technical art students. 

\noindent\textbf{\\2. Lack of Community Support.}
Many shared concerns about the scant support for art students learning to code in schools, including the lack of teaching assistants (I1), experienced peer students with technical backgrounds (P2, P4), and instructors' availability (P3).\\
\textbf{{$\bigstar$} Design Goals:} \n should be an easy-to-use and natural language-friendly chatbot embedded in prevalent creative coding or visualization editors like p5.js\footnote{\url{https://editor.p5js.org/}} or OpenProcessing\footnote{\url{https://openprocessing.org/sketch/create}} to forge constant support for students, which 1) offers suggestions, 2) addresses questions, 3) explains concepts and codes, 4) supports finding creative designs and connections, and 5) motivates students' code learning and design process.

\noindent\textbf{\\3. Mental Barriers with Coding.}
All participants expressed that the iterative debugging process of converting artistic ideas into code is always overwhelming and underestimated by the students given their limited technical experiences and insufficient support. \\
\textbf{{$\bigstar$} Design Goals:} \n should leverage LLMs to simplify the coding process: provide natural language examples and explanations, scaffold complicated tasks into approachable steps, and provide emotional backing, companionship, and validation. 

\noindent\textbf{\\4. Translation of Artistic Ideas into Code.}
Participants emphasize the difficulty of translating creative visions into code due to coding limitations. P2, P3, and P4 particularly resonate with this as they usually search and adapt online code templates to achieve their creative objectives. However, they find the available templates limited, particularly when extending to interaction and music.\\
\textbf{{$\bigstar$} Design Goals:} \n should incorporate design and various generative models to assist students in rendering visions as code: 1) use scaffolding to break down the art-to-code process into tangible steps, 2) use LLMs and image-to-text models to help students formulate their initial creative vision from scratch or existing designs, 3) use LLMs and text-to-image models to help students find creative connections, generate unique coding templates, and co-create visuals and interactive animations, 4) use text-to-music models to produce background sounds for artworks, and 5) use LLMs to finalize a creative vision, descriptions, and deployable code snippets for portfolios. 

\section{System Implementation \& Feasibility}
\n will be developed into a coding companion chatbot and functioned as an add-on for leading online creative coding and visualization editors. As a floating overlay on the browser, \n will use JavaScript for interfacing with the editors while leveraging generative model APIs and Python for backend operations, thus enabling real-time coding and design suggestions and supports. Besides, we will build a Flask app to test the workflow.

For a technical proof-of-concept, we break down the system into six stages and have already tested the feasibility. Users can interact with the overlaying chatbot about any of these stages for dedicated help, or they can navigate through a complete cycle. See Appendix for a detailed walkthrough and demo of technical feasibility:

\vspace{6px}
\noindent\textbf{Stage \#1: Project Ideation}
{\small
\\ Begin artwork conceptualization using textual or visual inputs.\\
\textbf{- Input}: Rough ideas, sketches, or online images.\\
\textbf{- Models}: Claude~\cite{Claude}, image-to-text (MiniGPT-4~\cite{MiniGPT4}, BLIP-2~\cite{BLIP2}).\\
\textbf{- Output}: Descriptive text and design suggestions for artwork.\vspace{6px}
}

\noindent\textbf{Stage \#2: Visual Design}
{\small
\\ Brainstorm artwork and generate coding templates.\\
\textbf{- Input}: Text description of artwork (from Stage \#1 or user).\\
\textbf{- Models}: Claude~\cite{Claude}, text-to-image (CLIP~\cite{CLIP}, VQGAN-CLIP~\cite{VQGAN}).\\
\textbf{- Output}: Visuals, creative connections, code templates, technical feasibility.\vspace{6px}
}

\noindent\textbf{Stage \#3: Audio Design}
{\small
\\ Suggest and craft background music for artwork.\\
\textbf{- Input}: Artwork design (from earlier stages or user).\\
\textbf{- Models}: Claude~\cite{Claude}, text-to-music (MusicLM~\cite{MusicLM}, {\sc MusicGen}~\cite{MusicGEN}, CLAP~\cite{CLAP}).\\
\textbf{- Output}: Music recommendations and sample audio files (.mp3).\vspace{6px}
}

\noindent\textbf{Stage \#4: Interaction Design}
{\small
\\ Develop user interactivity in sync with audio elements.\\
\textbf{- Input}: Artwork design (from earlier stages or user).\\
\textbf{- Models}: Claude~\cite{Claude}.\\
\textbf{- Output}: Coding editor updates with interactive feature templates.\vspace{6px}
}

\noindent\textbf{Stage 5 \& 6: Creative Vision \& Portfolio Delivery}
{\small
\\ Finalize creative visions, narratives, and coding for portfolio.\\
\textbf{- Input}: Artwork design (from earlier stages or user).\\
\textbf{- Models}: Claude~\cite{Claude}.\\
\textbf{- Output}: Polished vision, design narratives, code snippets (.html, .js, .css).\vspace{6px}
}

\noindent\textbf{Q\&A, Proactive Suggestions, Code Learning, Companionship}
{\small
\\ Analyze on-screen content, give suggestions, handle Q\&A, offer code explanations and learning guidance, and provide feedback for motivation.\\
\textbf{- Input}: Artwork design and on-screen materials.\\
\textbf{- Models}: Claude~\cite{Claude}, DOM Manipulation, Screen Capture API.\\
\textbf{- Output}: Chat interactions, screen highlights, and code explanations.
}

\section{Next Steps}
We will refine our stage and system designs by co-designing with 10 more non-technical art students. This involves prompt engineering, feature testing, and natural chat testing to optimize the interaction. Besides, we will implement sentimental understanding and expand the system's art contextual knowledge to make it more useful. After user testing, \n will be introduced to real portfolio studios, positioning it as a hands-on tool for teaching and learning.
\newpage
\appendix
\section*{Appendix: Prototype Links}
\begin{itemize}
    \item {\href{https://www.figma.com/file/JblNO3NXFIDs0XeIP1o2gk/PortfolioMentor?type=design&node-id=53-2&mode=design}{\faIcon{link} Design File}}
    \item {\href{https://www.figma.com/proto/JblNO3NXFIDs0XeIP1o2gk/PortfolioMentor?type=design&node-id=146-117&scaling=scale-down&page-id=53%3A2&starting-point-node-id=146%3A117}{\faIcon{link} Design Prototype}}
    \item {\href{https://drive.google.com/file/d/1JxENctu7uQV97vqynMI2RYE5TcBC1-j3/view?usp=sharing}{\faIcon{link} Design Video}}
\end{itemize}
\bibliographystyle{ACM-Reference-Format}
\bibliography{sample-base}


\begin{thebibliography}{23}


\ifx \showCODEN    \undefined \def \showCODEN     #1{\unskip}     \fi
\ifx \showDOI      \undefined \def \showDOI       #1{#1}\fi
\ifx \showISBNx    \undefined \def \showISBNx     #1{\unskip}     \fi
\ifx \showISBNxiii \undefined \def \showISBNxiii  #1{\unskip}     \fi
\ifx \showISSN     \undefined \def \showISSN      #1{\unskip}     \fi
\ifx \showLCCN     \undefined \def \showLCCN      #1{\unskip}     \fi
\ifx \shownote     \undefined \def \shownote      #1{#1}          \fi
\ifx \showarticletitle \undefined \def \showarticletitle #1{#1}   \fi
\ifx \showURL      \undefined \def \showURL       {\relax}        \fi
\providecommand\bibfield[2]{#2}
\providecommand\bibinfo[2]{#2}
\providecommand\natexlab[1]{#1}
\providecommand\showeprint[2][]{arXiv:#2}

\bibitem[Abrami and Barrett(2005)]%
        {multimediaportfolio}
\bibfield{author}{\bibinfo{person}{Philip Abrami} {and} \bibinfo{person}{Helen
  Barrett}.} \bibinfo{year}{2005}\natexlab{}.
\newblock \showarticletitle{Directions for Research and Development on
  Electronic Portfolios}.
\newblock \bibinfo{journal}{\emph{Canadian Journal of Learning and Technology /
  La revue canadienne de l’apprentissage et de la technologie}}
  \bibinfo{volume}{31}, \bibinfo{number}{3} (\bibinfo{date}{Oct}
  \bibinfo{year}{2005}).
\newblock
\showISSN{1499-6677}
\urldef\tempurl%
\url{https://www.learntechlib.org/p/43165/}
\showURL{%
\tempurl}


\bibitem[Agostinelli et~al\mbox{.}(2023)]%
        {MusicLM}
\bibfield{author}{\bibinfo{person}{Andrea Agostinelli},
  \bibinfo{person}{Timo~I. Denk}, \bibinfo{person}{Zalán Borsos},
  \bibinfo{person}{Jesse Engel}, \bibinfo{person}{Mauro Verzetti},
  \bibinfo{person}{Antoine Caillon}, \bibinfo{person}{Qingqing Huang},
  \bibinfo{person}{Aren Jansen}, \bibinfo{person}{Adam Roberts},
  \bibinfo{person}{Marco Tagliasacchi}, \bibinfo{person}{Matt Sharifi},
  \bibinfo{person}{Neil Zeghidour}, {and} \bibinfo{person}{Christian Frank}.}
  \bibinfo{year}{2023}\natexlab{}.
\newblock \showarticletitle{MusicLM: Generating Music From Text}.
\newblock  \bibinfo{number}{arXiv:2301.11325} (\bibinfo{date}{Jan}
  \bibinfo{year}{2023}).
\newblock
\urldef\tempurl%
\url{https://doi.org/10.48550/arXiv.2301.11325}
\showDOI{\tempurl}
\newblock
\shownote{arXiv:2301.11325 [cs, eess]}.


\bibitem[Angert et~al\mbox{.}(2023)]%
        {creativecodingexploration}
\bibfield{author}{\bibinfo{person}{Tyler Angert},
  \bibinfo{person}{Miroslav~Ivan Suzara}, \bibinfo{person}{Jenny Han},
  \bibinfo{person}{Christopher~Lawrence Pondoc}, {and}
  \bibinfo{person}{Hariharan Subramonyam}.} \bibinfo{year}{2023}\natexlab{}.
\newblock \showarticletitle{Spellburst: A Node-based Interface for Exploratory
  Creative Coding with Natural Language Prompts}.
\newblock  (\bibinfo{date}{Aug} \bibinfo{year}{2023}).
\newblock
\urldef\tempurl%
\url{https://doi.org/10.1145/3586183.3606719}
\showDOI{\tempurl}
\newblock
\shownote{arXiv:2308.03921 [cs]}.


\bibitem[Anthropic({[n.\,d.]})]%
        {Claude}
\bibfield{author}{\bibinfo{person}{Anthropic}.}
  \bibinfo{year}{[n.\,d.]}\natexlab{}.
\newblock \bibinfo{title}{Claude API Access}.
\newblock
\newblock
\urldef\tempurl%
\url{https://www.anthropic.com/earlyaccess}
\showURL{%
\tempurl}


\bibitem[Biermann et~al\mbox{.}(2022)]%
        {companionshipcreative}
\bibfield{author}{\bibinfo{person}{Oloff~C. Biermann}, \bibinfo{person}{Ning~F.
  Ma}, {and} \bibinfo{person}{Dongwook Yoon}.} \bibinfo{year}{2022}\natexlab{}.
\newblock \showarticletitle{From Tool to Companion: Storywriters Want AI
  Writers to Respect Their Personal Values and Writing Strategies}. In
  \bibinfo{booktitle}{\emph{Proceedings of the 2022 ACM Designing Interactive
  Systems Conference}} \emph{(\bibinfo{series}{DIS ’22})}.
  \bibinfo{publisher}{Association for Computing Machinery},
  \bibinfo{address}{New York, NY, USA}, \bibinfo{pages}{1209–1227}.
\newblock
\showISBNx{978-1-4503-9358-4}
\urldef\tempurl%
\url{https://doi.org/10.1145/3532106.3533506}
\showDOI{\tempurl}


\bibitem[Copet et~al\mbox{.}(2023)]%
        {MusicGEN}
\bibfield{author}{\bibinfo{person}{Jade Copet}, \bibinfo{person}{Felix Kreuk},
  \bibinfo{person}{Itai Gat}, \bibinfo{person}{Tal Remez},
  \bibinfo{person}{David Kant}, \bibinfo{person}{Gabriel Synnaeve},
  \bibinfo{person}{Yossi Adi}, {and} \bibinfo{person}{Alexandre Défossez}.}
  \bibinfo{year}{2023}\natexlab{}.
\newblock \showarticletitle{Simple and Controllable Music Generation}.
\newblock  \bibinfo{number}{arXiv:2306.05284} (\bibinfo{date}{Jun}
  \bibinfo{year}{2023}).
\newblock
\urldef\tempurl%
\url{http://arxiv.org/abs/2306.05284}
\showURL{%
\tempurl}
\newblock
\shownote{arXiv:2306.05284 [cs, eess]}.


\bibitem[Crowson et~al\mbox{.}(2022)]%
        {VQGAN}
\bibfield{author}{\bibinfo{person}{Katherine Crowson}, \bibinfo{person}{Stella
  Biderman}, \bibinfo{person}{Daniel Kornis}, \bibinfo{person}{Dashiell
  Stander}, \bibinfo{person}{Eric Hallahan}, \bibinfo{person}{Louis
  Castricato}, {and} \bibinfo{person}{Edward Raff}.}
  \bibinfo{year}{2022}\natexlab{}.
\newblock \showarticletitle{VQGAN-CLIP: Open Domain Image Generation and
  Editing with Natural Language Guidance}.
\newblock  \bibinfo{number}{arXiv:2204.08583} (\bibinfo{date}{Sep}
  \bibinfo{year}{2022}).
\newblock
\urldef\tempurl%
\url{http://arxiv.org/abs/2204.08583}
\showURL{%
\tempurl}
\newblock
\shownote{arXiv:2204.08583 [cs]}.


\bibitem[Gero et~al\mbox{.}(2023)]%
        {socialDynamics}
\bibfield{author}{\bibinfo{person}{Katy~Ilonka Gero}, \bibinfo{person}{Tao
  Long}, {and} \bibinfo{person}{Lydia~B. Chilton}.}
  \bibinfo{year}{2023}\natexlab{}.
\newblock \showarticletitle{Social dynamics of AI support in creative writing}.
  In \bibinfo{booktitle}{\emph{Proceedings of the 2023 CHI Conference on Human
  Factors in Computing Systems}}. \bibinfo{pages}{1–15}.
\newblock


\bibitem[Kim et~al\mbox{.}(2022)]%
        {stylette}
\bibfield{author}{\bibinfo{person}{Tae~Soo Kim}, \bibinfo{person}{DaEun Choi},
  \bibinfo{person}{Yoonseo Choi}, {and} \bibinfo{person}{Juho Kim}.}
  \bibinfo{year}{2022}\natexlab{}.
\newblock \showarticletitle{Stylette: Styling the Web with Natural Language}.
  In \bibinfo{booktitle}{\emph{Proceedings of the 2022 CHI Conference on Human
  Factors in Computing Systems}} (New Orleans, LA, USA)
  \emph{(\bibinfo{series}{CHI '22})}. \bibinfo{publisher}{Association for
  Computing Machinery}, \bibinfo{address}{New York, NY, USA}, Article
  \bibinfo{articleno}{5}, \bibinfo{numpages}{17}~pages.
\newblock
\showISBNx{9781450391573}
\urldef\tempurl%
\url{https://doi.org/10.1145/3491102.3501931}
\showDOI{\tempurl}


\bibitem[Leitner and Subasi(2016)]%
        {creativevision}
\bibfield{author}{\bibinfo{person}{Michael Leitner} {and}
  \bibinfo{person}{Özge Subasi}.} \bibinfo{year}{2016}\natexlab{}.
\newblock \showarticletitle{Arty Portfolios: Manifesting Artistic Work in
  Interaction Design Research}. In \bibinfo{booktitle}{\emph{Proceedings of the
  9th Nordic Conference on Human-Computer Interaction}}.
  \bibinfo{publisher}{ACM}, \bibinfo{address}{Gothenburg Sweden},
  \bibinfo{pages}{1–10}.
\newblock
\showISBNx{978-1-4503-4763-1}
\urldef\tempurl%
\url{https://doi.org/10.1145/2971485.2971515}
\showDOI{\tempurl}


\bibitem[Li et~al\mbox{.}(2023)]%
        {BLIP2}
\bibfield{author}{\bibinfo{person}{Junnan Li}, \bibinfo{person}{Dongxu Li},
  \bibinfo{person}{Silvio Savarese}, {and} \bibinfo{person}{Steven Hoi}.}
  \bibinfo{year}{2023}\natexlab{}.
\newblock \showarticletitle{BLIP-2: Bootstrapping Language-Image Pre-training
  with Frozen Image Encoders and Large Language Models}.
\newblock  \bibinfo{number}{arXiv:2301.12597} (\bibinfo{date}{Jun}
  \bibinfo{year}{2023}).
\newblock
\urldef\tempurl%
\url{https://doi.org/10.48550/arXiv.2301.12597}
\showDOI{\tempurl}
\newblock
\shownote{arXiv:2301.12597 [cs]}.


\bibitem[Liu et~al\mbox{.}(2023)]%
        {disco}
\bibfield{author}{\bibinfo{person}{Vivian Liu}, \bibinfo{person}{Tao Long},
  \bibinfo{person}{Nathan Raw}, {and} \bibinfo{person}{Lydia Chilton}.}
  \bibinfo{year}{2023}\natexlab{}.
\newblock \showarticletitle{Generative Disco: Text-to-Video Generation for
  Music Visualization}.
\newblock  \bibinfo{number}{arXiv:2304.08551} (\bibinfo{date}{Apr}
  \bibinfo{year}{2023}).
\newblock
\urldef\tempurl%
\url{https://doi.org/10.48550/arXiv.2304.08551}
\showDOI{\tempurl}
\newblock
\shownote{arXiv:2304.08551 [cs]}.


\bibitem[Long et~al\mbox{.}(2023)]%
        {Tweetorial_hook}
\bibfield{author}{\bibinfo{person}{Tao Long}, \bibinfo{person}{Dorothy Zhang},
  \bibinfo{person}{Grace Li}, \bibinfo{person}{Batool Taraif},
  \bibinfo{person}{Samia Menon}, \bibinfo{person}{Kynnedy~Simone Smith},
  \bibinfo{person}{Sitong Wang}, \bibinfo{person}{Katy~Ilonka Gero}, {and}
  \bibinfo{person}{Lydia~B. Chilton}.} \bibinfo{year}{2023}\natexlab{}.
\newblock \showarticletitle{Tweetorial Hooks: Generative AI Tools to Motivate
  Science on Social Media}. In \bibinfo{booktitle}{\emph{Proceedings of the
  14th Conference on Computational Creativity}} \emph{(\bibinfo{series}{ICCC
  '23})}. Association for Computational Creativity.
\newblock


\bibitem[Louie et~al\mbox{.}(2020)]%
        {NoviceAImusic}
\bibfield{author}{\bibinfo{person}{Ryan Louie}, \bibinfo{person}{Andy Coenen},
  \bibinfo{person}{Cheng~Zhi Huang}, \bibinfo{person}{Michael Terry}, {and}
  \bibinfo{person}{Carrie~J. Cai}.} \bibinfo{year}{2020}\natexlab{}.
\newblock \bibinfo{booktitle}{\emph{Novice-AI Music Co-Creation via AI-Steering
  Tools for Deep Generative Models}}.
\newblock \bibinfo{publisher}{Association for Computing Machinery},
  \bibinfo{address}{New York, NY, USA}, \bibinfo{pages}{1–13}.
\newblock
\showISBNx{9781450367080}
\urldef\tempurl%
\url{https://doi.org/10.1145/3313831.3376739}
\showURL{%
\tempurl}


\bibitem[MacNeil et~al\mbox{.}(2021)]%
        {scaffolding}
\bibfield{author}{\bibinfo{person}{Stephen MacNeil}, \bibinfo{person}{Zijian
  Ding}, \bibinfo{person}{Kexin Quan}, \bibinfo{person}{Thomas~j Parashos},
  \bibinfo{person}{Yajie Sun}, {and} \bibinfo{person}{Steven~P. Dow}.}
  \bibinfo{year}{2021}\natexlab{}.
\newblock \showarticletitle{Framing Creative Work: Helping Novices Frame Better
  Problems through Interactive Scaffolding}. In
  \bibinfo{booktitle}{\emph{Creativity and Cognition}} (Virtual Event, Italy).
  \bibinfo{publisher}{Association for Computing Machinery},
  \bibinfo{address}{New York, NY, USA}, Article \bibinfo{articleno}{30},
  \bibinfo{numpages}{10}~pages.
\newblock
\showISBNx{9781450383769}
\urldef\tempurl%
\url{https://doi.org/10.1145/3450741.3465261}
\showDOI{\tempurl}


\bibitem[Mcnutt et~al\mbox{.}(2023)]%
        {classroomCHIcreativecoding}
\bibfield{author}{\bibinfo{person}{Andrew~M Mcnutt}, \bibinfo{person}{Anton
  Outkine}, {and} \bibinfo{person}{Ravi Chugh}.}
  \bibinfo{year}{2023}\natexlab{}.
\newblock \showarticletitle{A Study of Editor Features in a Creative Coding
  Classroom}. In \bibinfo{booktitle}{\emph{Proceedings of the 2023 CHI
  Conference on Human Factors in Computing Systems}}. \bibinfo{publisher}{ACM},
  \bibinfo{address}{Hamburg Germany}, \bibinfo{pages}{1–15}.
\newblock
\showISBNx{978-1-4503-9421-5}
\urldef\tempurl%
\url{https://doi.org/10.1145/3544548.3580683}
\showDOI{\tempurl}


\bibitem[Mitchell and Bown(2013)]%
        {creativecoderneeds}
\bibfield{author}{\bibinfo{person}{Mark~C. Mitchell} {and}
  \bibinfo{person}{Oliver Bown}.} \bibinfo{year}{2013}\natexlab{}.
\newblock \showarticletitle{Towards a creativity support tool in processing:
  understanding the needs of creative coders}. In
  \bibinfo{booktitle}{\emph{Proceedings of the 25th Australian Computer-Human
  Interaction Conference: Augmentation, Application, Innovation,
  Collaboration}} \emph{(\bibinfo{series}{OzCHI ’13})}.
  \bibinfo{publisher}{Association for Computing Machinery},
  \bibinfo{address}{New York, NY, USA}, \bibinfo{pages}{143–146}.
\newblock
\showISBNx{978-1-4503-2525-7}
\urldef\tempurl%
\url{https://doi.org/10.1145/2541016.2541096}
\showDOI{\tempurl}


\bibitem[Petridis et~al\mbox{.}(2023)]%
        {promptinfuser}
\bibfield{author}{\bibinfo{person}{Savvas Petridis}, \bibinfo{person}{Michael
  Terry}, {and} \bibinfo{person}{Carrie~Jun Cai}.}
  \bibinfo{year}{2023}\natexlab{}.
\newblock \showarticletitle{PromptInfuser: Bringing User Interface Mock-Ups to
  Life with Large Language Models}. In \bibinfo{booktitle}{\emph{Extended
  Abstracts of the 2023 CHI Conference on Human Factors in Computing Systems}}
  (Hamburg, Germany) \emph{(\bibinfo{series}{CHI EA '23})}.
  \bibinfo{publisher}{Association for Computing Machinery},
  \bibinfo{address}{New York, NY, USA}, Article \bibinfo{articleno}{237},
  \bibinfo{numpages}{6}~pages.
\newblock
\showISBNx{9781450394222}
\urldef\tempurl%
\url{https://doi.org/10.1145/3544549.3585628}
\showDOI{\tempurl}


\bibitem[Radford et~al\mbox{.}(2021)]%
        {CLIP}
\bibfield{author}{\bibinfo{person}{Alec Radford}, \bibinfo{person}{Jong~Wook
  Kim}, \bibinfo{person}{Chris Hallacy}, \bibinfo{person}{Aditya Ramesh},
  \bibinfo{person}{Gabriel Goh}, \bibinfo{person}{Sandhini Agarwal},
  \bibinfo{person}{Girish Sastry}, \bibinfo{person}{Amanda Askell},
  \bibinfo{person}{Pamela Mishkin}, \bibinfo{person}{Jack Clark},
  \bibinfo{person}{Gretchen Krueger}, {and} \bibinfo{person}{Ilya Sutskever}.}
  \bibinfo{year}{2021}\natexlab{}.
\newblock \showarticletitle{Learning Transferable Visual Models From Natural
  Language Supervision}.
\newblock  \bibinfo{number}{arXiv:2103.00020} (\bibinfo{date}{Feb}
  \bibinfo{year}{2021}).
\newblock
\urldef\tempurl%
\url{https://doi.org/10.48550/arXiv.2103.00020}
\showDOI{\tempurl}
\newblock
\shownote{arXiv:2103.00020 [cs]}.


\bibitem[Sarkar et~al\mbox{.}(2022)]%
        {companionshipcoding}
\bibfield{author}{\bibinfo{person}{Advait Sarkar}, \bibinfo{person}{Andrew~D.
  Gordon}, \bibinfo{person}{Carina Negreanu}, \bibinfo{person}{Christian
  Poelitz}, \bibinfo{person}{Sruti~Srinivasa Ragavan}, {and}
  \bibinfo{person}{Ben Zorn}.} \bibinfo{year}{2022}\natexlab{}.
\newblock \showarticletitle{What is it like to program with artificial
  intelligence?}
\newblock  \bibinfo{number}{arXiv:2208.06213} (\bibinfo{date}{Oct}
  \bibinfo{year}{2022}).
\newblock
\urldef\tempurl%
\url{https://doi.org/10.48550/arXiv.2208.06213}
\showDOI{\tempurl}
\newblock
\shownote{arXiv:2208.06213 [cs]}.


\bibitem[Wang et~al\mbox{.}(2023)]%
        {reelframer}
\bibfield{author}{\bibinfo{person}{Sitong Wang}, \bibinfo{person}{Samia Menon},
  \bibinfo{person}{Tao Long}, \bibinfo{person}{Keren Henderson},
  \bibinfo{person}{Dingzeyu Li}, \bibinfo{person}{Kevin Crowston},
  \bibinfo{person}{Mark Hansen}, \bibinfo{person}{Jeffrey~V. Nickerson}, {and}
  \bibinfo{person}{Lydia~B. Chilton}.} \bibinfo{year}{2023}\natexlab{}.
\newblock \showarticletitle{ReelFramer: Co-creating News Reels on Social Media
  with Generative AI}.
\newblock  \bibinfo{number}{arXiv:2304.09653} (\bibinfo{date}{Apr}
  \bibinfo{year}{2023}).
\newblock
\urldef\tempurl%
\url{https://doi.org/10.48550/arXiv.2304.09653}
\showDOI{\tempurl}
\newblock
\shownote{arXiv:2304.09653 [cs]}.


\bibitem[Wu et~al\mbox{.}(2023)]%
        {CLAP}
\bibfield{author}{\bibinfo{person}{Yusong Wu}, \bibinfo{person}{Ke Chen},
  \bibinfo{person}{Tianyu Zhang}, \bibinfo{person}{Yuchen Hui},
  \bibinfo{person}{Taylor Berg-Kirkpatrick}, {and} \bibinfo{person}{Shlomo
  Dubnov}.} \bibinfo{year}{2023}\natexlab{}.
\newblock \showarticletitle{Large-scale Contrastive Language-Audio Pretraining
  with Feature Fusion and Keyword-to-Caption Augmentation}.
\newblock  \bibinfo{number}{arXiv:2211.06687} (\bibinfo{date}{Apr}
  \bibinfo{year}{2023}).
\newblock
\urldef\tempurl%
\url{http://arxiv.org/abs/2211.06687}
\showURL{%
\tempurl}
\newblock
\shownote{arXiv:2211.06687 [cs, eess]}.


\bibitem[Zhu et~al\mbox{.}(2023)]%
        {MiniGPT4}
\bibfield{author}{\bibinfo{person}{Deyao Zhu}, \bibinfo{person}{Jun Chen},
  \bibinfo{person}{Xiaoqian Shen}, \bibinfo{person}{Xiang Li}, {and}
  \bibinfo{person}{Mohamed Elhoseiny}.} \bibinfo{year}{2023}\natexlab{}.
\newblock \showarticletitle{MiniGPT-4: Enhancing Vision-Language Understanding
  with Advanced Large Language Models}.
\newblock \bibinfo{journal}{\emph{arXiv preprint arXiv:2304.10592}}
  (\bibinfo{year}{2023}).
\newblock


\end{thebibliography}

\end{document}